\documentclass[twocolumn,amsmath,amssymb]{revtex4}

\usepackage{graphicx}
\usepackage{dcolumn}
\usepackage{bm}

\def\Ho{$H_\circ$}
\def\omegao{$\Omega_\circ$}
\def\to{$t_\circ$}

\begin{document}


\title{Note on the classical solutions of Friedmann's equation}

\author{Arthur Viglioni}
\email{arthurviglioni@yahoo.com.br}
\author{Domingos Soares}
\email{dsoares@fisica.ufmg.br}


\affiliation{Physics Department, Universidade Federal de Minas Gerais,\\
Belo Horizonte, MG, Brazil}

\date{\today}

\begin{abstract}
Graphical representations of classical Friedmann's models are  
often misleading when one considers the age of the universe. 
Most textbooks disregard conceptual differences in the 
representations, as far as ages are concerned. 
We discuss the details of the scale-factor versus 
time function for Friedmann's solutions in the time range that 
includes the ages of model universes.  
\end{abstract}



\maketitle

\section{Introduction}

Modern Big Bang cosmological models are described by modified 
Friedmann's models, with the inclusion of a cosmological constant 
(e.g., \cite{rind}, p. 403). Hence there is always 
renewed interest in all aspects of Friedmann's models. 

Classical Friedmann's  models are fully described by the expansion 
rate of the scale factor R(t), \Ho=$\dot{\rm R}$/R --- the so-called 
\emph{Hubble parameter}, evaluated \emph{now} (t=\to, the age of 
a given model) ---, and the density parameter 
\omegao$=\rho_\circ/\rho_{c\circ}$, where $\rho_\circ$ is 
the model mass density and $\rho_{c\circ}=3H_\circ^2/8\pi G$ 
is the critical density --- the density of the flat model ---, both on t=\to.

The classical Friedmann's equation, i.e.,  without the cosmological 
constant, is written as (see, e.g., \cite{caos}, chap. 27 and 
\cite{desou}, chap. 2):

\begin{equation}
\label{eq:fried}
\left(\frac{dR}{dt}\right)^2 - \frac{H_\circ^2\Omega_\circ}{R} = 
-H_\circ^2(\Omega_\circ - 1), 
\end{equation}

\noindent where R(\to) is, conventionally, set to unity. 
The solution for the flat model is readily obtained inserting 
$\Omega_\circ=1$ in eq.~\ref{eq:fried}:

\begin{equation}
\label{eq:flat}
R(t) = \left(\frac{t}{t_\circ}\right)^{2/3}, 
\end{equation}

\noindent with \to=2/(3\Ho) being the flat model's age. For the closed 
model ($\Omega_\circ>1$) the solution is expressed in the 
parametric form (see \cite{caos}, eqs. 27.24 and 27.26):

\begin{equation}
\label{eq:closed1}
R(x)=\frac{1}{2}\frac{\Omega_\circ}{\Omega_\circ-1}\left[1-\cos(x)\right], 
\end{equation}
\begin{equation}
\label{eq:closed2}
t(x)=\frac{1}{2H_\circ}\frac{\Omega_\circ}{(\Omega_\circ-1)^{3/2}}\left[x-\sin(x)\right], 
\end{equation}

\noindent where the parameter is $x\ge0$. Likewise, the solution for 
the open model ($\Omega_\circ<1$) is given by:

\begin{equation}
\label{eq:open1}
R(x)=\frac{1}{2}\frac{\Omega_\circ}{1-\Omega_\circ}\left[\cosh(x)-1\right], 
\end{equation}
\begin{equation}
\label{eq:open2}
t(x)=\frac{1}{2H_\circ}\frac{\Omega_\circ}{(1-\Omega_\circ)^{3/2}}\left[\sinh(x)-x\right]. 
\end{equation}

\begin{figure}
\includegraphics[width=8cm]{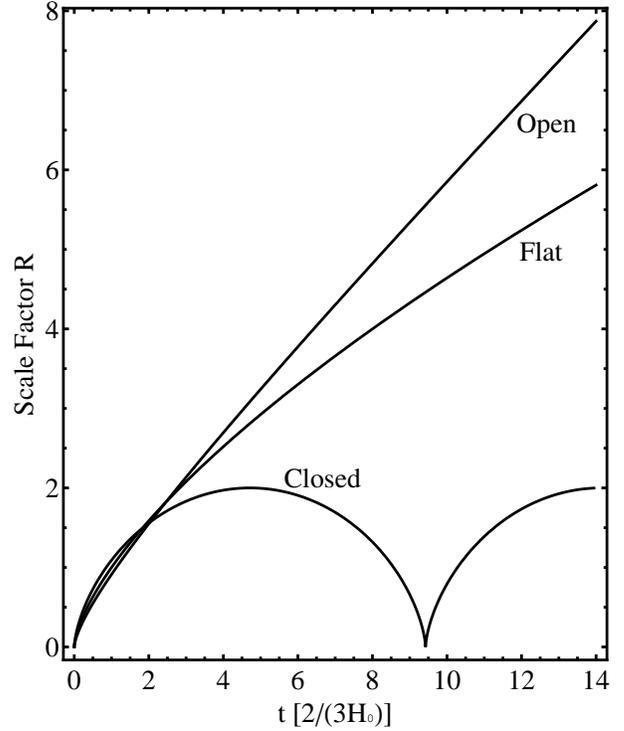}
\caption{\label{fig:fried1} Open (\omegao=0.5), flat and closed
(\omegao=2) Friedmann's models. This is an example of their most 
common graphical representation. Time, in the 
abscissas, is in units of 2/(3\Ho),  the age of the flat model. 
The fine features at $R<2$ are absent in most textbooks and 
popular reviews of cosmology.
}
\end{figure}

These solutions for the open (\omegao=0.5), flat or critical (\omegao=1) and 
closed (\omegao=2) models are depicted in Fig.~\ref{fig:fried1}. It is 
worthwhile stressing that the subsequent discussion does not depend on 
the $\Omega_\circ$ values chosen for the close and open models.

Most textbooks, or qualitative papers on modern cosmology, present 
diagrams like this one, when describing Friedmann's models. With rare 
exceptions, they differ from our Fig.~\ref{fig:fried1} in two main features. 
First, there is no quantitative axes, which are generically labeled "scale  
factor'' and "time''. Second, the fine features at the small R(t) range 
are not considered, let alone plotted (see R(t)$<2$ in 
Fig.~\ref{fig:fried1}). Such differences lead to a wrong 
conceptual apprehension of the models, concerning their ages, i.e., the 
times corresponding to R=1.
 
In the next section, we show why it is important to present a precise 
graphical representation of Friedmann's models, in the light of the 
models' ages. The last section concludes with a quantitative 
account on the relative differences between classical Friedmann's models  
as functions of cosmic time.

\section{Ages of Friedmann's universes}

\begin{figure}
\includegraphics[width=6cm]{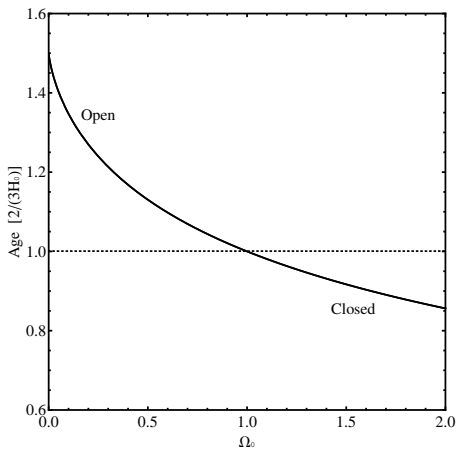}
\caption{\label{fig:age} The ages of open, flat and closed 
Friedmann's models as functions of the density parameter \omegao. 
Note that the ages of open models are always greater than the ages 
of flat (\omegao=1) and closed models. The vertical axis is in units 
of the flat model age, namely, 2/(3\Ho).   
}
\end{figure}

The ages of Friedmann's universes are obtained by making $R=1$ in 
eqs.~\ref{eq:flat}, \ref{eq:closed1} and \ref{eq:open1}. Then, with the aid 
of eqs.~\ref{eq:closed2} and \ref{eq:open2}, one gets the ages of the open, 
flat and closed models as functions of the density parameter \omegao. The 
result is plotted in Fig.~\ref{fig:age}. 

The age of the open model is the greatest amongst Friedmann's 
classical solutions. This is not clear when one examines 
Fig.~\ref{fig:fried1}, because the relevant range of the scale  
factor R(t)$\approx 1$ is not immediately apparent in the figure. This 
sort of graphical representation is predominant in most textbooks. 
For example, it is seen in Harrison \cite{harr}, Fig. 18.6, p. 360, 
Carroll \& Ostlie \cite{caos}, Fig. 27.4, p. 1230, Rindler 
\cite{rind}, Fig. 18.2, p. 402, Shu \cite{shu}, Fig. 15.7, p. 362 
and Box 15.4, p. 368. Harrison discusses ages after 
presenting Fig. 18.6. He calls the reader's attention to their 
differences, but the conclusion remains conceptually 
inconsistent with the diagram shown in Fig. 18.6. In Carroll \& Ostlie 
there is a hint of fine features in the small R(t) range.  
In fact, they comply with both desired features mentioned in the 
introductory section, but do not discuss the age-related issue.
Nevertheless, they constitute a rare exception, in the cosmological 
literature, when plotting solutions of Friedmann's equations. 

\begin{figure}
\includegraphics[width=8cm]{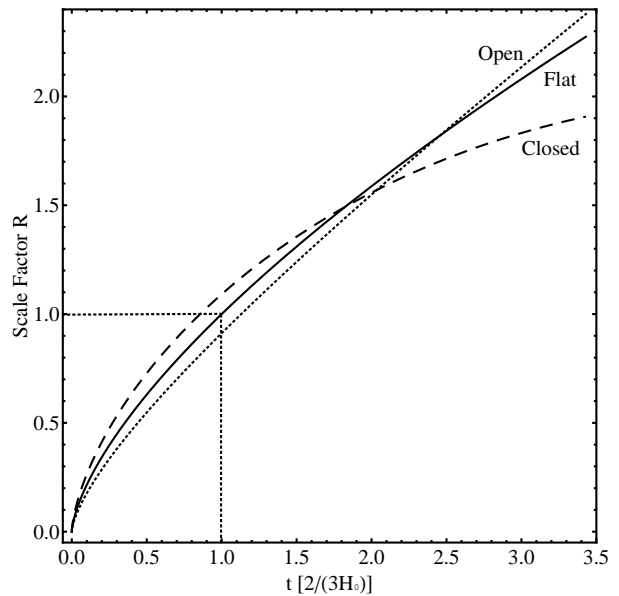}
\caption{\label{fig:fried2} Open (\omegao=0.5), flat and closed 
(\omegao=2) Friedmann's models are now shown in the range that 
includes R(\to)=1, right in the middle of the vertical axis. The age of 
the flat model is shown in the diagram, \to = 9 Gyr, for \Ho = 
72 km/s Mpc$^{-1}$ \cite{freed}. Note that, according to eq.~\ref{eq:fried} 
and irrespective of \omegao , the three curves share the 
same slope $dR/dt$ at $R=1$ (i.e., \emph{now}), for each model, and 
consequently, the same Hubble parameter $\dot{R}/R=H_\circ$.
}
\end{figure}

The age of the universe is assigned to R(\to)=1. In Fig.~\ref{fig:fried2},  
we plot R(t) in the range appropriate to show the ages of the models. 
Contrary to the representation in Fig.~\ref{fig:fried1}, 
it is now clear that the closed model has the largest R(t), and the smallest 
age. Around t=2, the closed model performs two crossovers, 
first with the flat model and then with the open model. Notice 
also a third crossover, shortly later, between the open and flat models. 
Following the crossovers, the behavior of R(t) is the usual one, 
as shown in Fig.~\ref{fig:fried1}. 

If one does not carefully examine the numerical scale of  
the vertical axis in Fig.~\ref{fig:fried1}, conclusions about the ages  
of the models are confusing. Things may get worse because, usually, 
as mentioned above, the plot is shown without a numerical scale 
in both axes.  

\section{Closed and open models versus the flat model}

In the very small R(t), the term on the right side of Friedmann's 
equation becomes negligible, when compared to the second term 
on the left side. This holds in the range 
$R(t)\ll \Omega_\circ/\left|\Omega_\circ-1\right|$, i.e., 
$R(t)\ll 1$, for the models studied here. The approximate solution  
is given by 

\begin{figure}
\bigskip
\includegraphics[width=8cm]{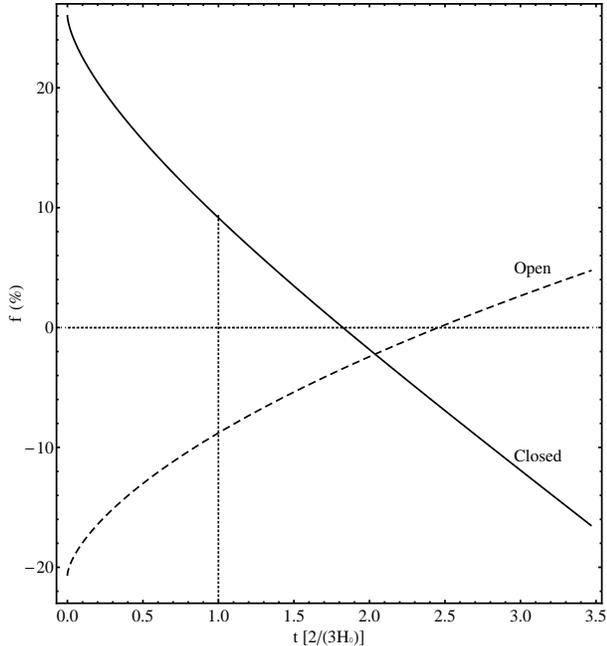}
\caption{\label{fig:diff} The relative differences between the closed 
and open models and the flat model as a function of time. The differences 
at $t\rightarrow 0$ are easily calculated from eq.~\ref{eq:fried0}. 
Note the crossover times with the flat model, 
marked by the horizontal dotted line. The vertical dotted line points 
to the age of the flat model (t=1).
}
\end{figure}
\begin{equation}
\label{eq:fried0}
R(t) \cong \Omega_\circ^{1/3}\left(\frac{t}{t_\circ}\right)^{2/3} =  
\Omega_\circ^{1/3}R_{flat}(t), 
\end{equation}

\noindent where \to=2/(3\Ho) is the flat model's age. Such an 
approximation confirms the fact that, for $t\rightarrow 0$, R(t) 
is larger for the closed model, as plotted in Fig.~\ref{fig:fried2}.

The relative differences among the models may be investigated 
by the percentage function $f=100\times(R - R_{flat})/R_{flat}$.
With the help of eq.~\ref{eq:fried0}, one gets 
$f=100\times(\Omega_\circ^{1/3}-1)$, which means 
$f=+26$\%, for the closed model,  
and $f=-21$\%, for the open model, at $t\approx 0$. 
This clearly shows that the models are quite different early on 
in the cosmic history, being the relative differences larger 
than around t=\to\ ($\left| f\right| \approx 9$\%, for the closed and open models). 

Fig.~\ref{fig:diff} shows the exact function $f(t)$. It is 
worthwhile noticing the location of the crossover times mentioned 
in section II, and that they occur after the ages of all models.

\bigskip

\section{Conclusions}

\begin{figure}
\includegraphics[width=8cm]{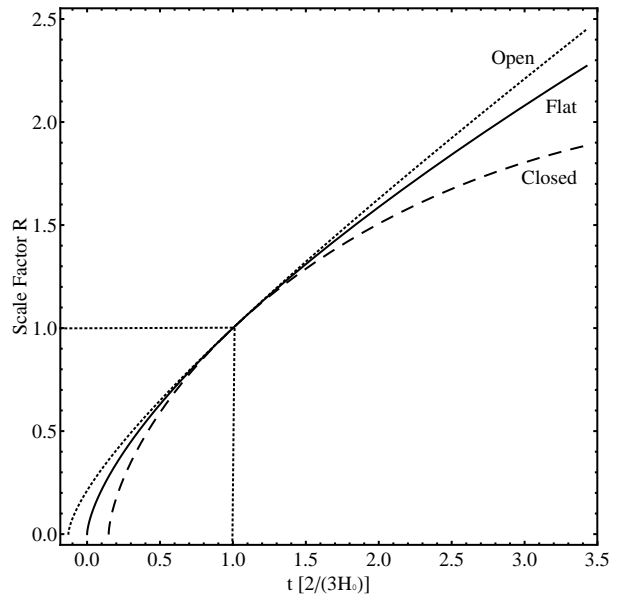}
\caption{\label{fig:fried3} Open (\omegao=0.5), flat and closed 
(\omegao=2) Friedmann's models have the same slope --- or Hubble's 
constant --- at $R=1$ (\emph{now}). Different ages \to\ appear along 
the time axis: \to$=1.0_{-0.14}^{+0.13}$ for the closed ($-$) and open ($+$) 
models (see also Fig.~\ref{fig:age}). 
}
\end{figure}

We have shown that the ages of Friedmann's classical universes are 
better appreciated when the fine features of the scale-factor 
function R(t), present in the range of times from $t=0$ to $t\approx 
3\times  2/(3H_\circ)$, are represented in detail (see 
Fig.~\ref{fig:fried2}). This is an alternative way to that adopted 
by Harrison (\cite{harr}, Fig. 18.7, p. 360) and Linder (\cite{lind}, 
Fig. 2.3, p. 32). These authors choose to stress the fact that all 
models have the same Hubble's parameter \emph{now} ($R=1$), as remarked in 
the legend of our Fig.~\ref{fig:fried2}. Those two figures are equivalent 
to our Fig.~\ref{fig:fried2} just by sliding the closed model curve forwards 
and the open model backwards, along the time axis, until they touch the 
flat model curve at $t=1$ (see Fig.~\ref{fig:fried3}).  

We calculate the relative differences between the closed 
and open models and the flat model, as a function of time 
(Fig.~\ref{fig:diff}). The models are quantitatively different 
right from the beginning, pass through crossovers around 
$t\approx  2\times 2/(3H_\circ)$ before diverging for $t\gg 2/(3H_\circ)$.

\begin{acknowledgments}
We wish to thank helpful discussion with Maxwell Rosa, Nat\'alia
M\'oller and Osvaldo Assun\c c\~ao.  
\end{acknowledgments}

\end{document}